\documentclass[aps,prb,twocolumn,floatfix]{revtex4}
\usepackage{epsfig,amsmath,amssymb,color}
\begin{document}

\title{Reducing entanglement with symmetries: application to persistent currents in impurity problems}
\author{A. E. Feiguin}
\affiliation{Department of Physics and Astronomy, University of Wyoming, Laramie, Wyoming 82071, USA}
\author{C. A. B\"usser}
\affiliation{Department of Physics and Astronomy, University of Wyoming, Laramie, Wyoming 82071, USA}

\date{\today}
\begin{abstract}
We show how canonical transformations can map problems with impurities coupled to non-interacting rings onto a similar problem with open boundary conditions. The consequent reduction of entanglement, and the fact the density matrix renormalization group (DMRG) is optimally suited for open boundary conditions, increases the efficiency of the method exponentially, making it an unprecedented tool to study persistent currents. We demonstrate its application to the case of the one-channel and two-channel Kondo problems, finding interesting connections between the two.
\end{abstract}

\maketitle

\section{Introduction}

The transport and non-equilibrium behavior of strongly correlated quantum many-body systems is one of the most challenging problems in condensed matter physics. 
Interactions can give rise some complex and intriguing phenomena,
such as the Kondo effect, with counter-intuitive transport properties in systems with impurities.\cite{Hewson}
Thanks to advances in nanofabrication, experimentalist
can routinely manufacture nanostructures that resemble artificial atoms
--quantum dots--
that can be manipulated with an extreme degree of control
\cite{Hanson2007,Goldhaber-GordonReview}, realizing an ideal playground to test transport theories. 
Understanding the physical phenomena arising in these systems has a fundamental technological interest, since it could lead to the development of the next generation of electronic devices.

The study of persistent currents in mesoscopic systems dates back to the 1980's \cite{imry}. 
A particular problem that has attracted a great deal of interest is the behavior of persistent currents through quantum-dot devices\cite{carlos}.
They can shed light on the Fermi liquid properties of these systems \cite{affleck-sorensen-2006}, and provide a sorely needed tool to calculate conductance \cite{rejec,armin}, and other transport properties \cite{shastry,doug,rodrigo}. 
Interferometry devices can also serve as probes for fractional statistics \cite{chetan}, and spin-charge separation \cite{julian}. 
Unfortunately, the lack of well controlled analytical methods that can deal with Kondo physics, especially in a ring geometry, can be a source for disagreement between different theoretical treatments \cite{cho2001,eckle,aligia,sorensen2005}.
Since quantum Monte Carlo \cite{qmc} and Numerical Renormalization Group (NRG) \cite{bulla} cannot be applied to the case of a magnetic flux threading the ring (appearing as a complex phase in the hopping), it would be highly desirable to count with a reliable numerical technique to treat this problem. 

\begin{figure}
\epsfig {file=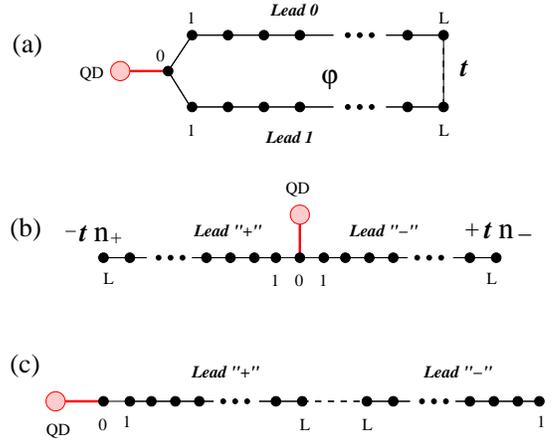,width=80mm}
\caption{
Cartoons showing the impurity or quantum-dot (QD) coupled to a ring (a) and the equivalent systems with open boundary conditions: (b) with the impurity in the center, and (c) at the edge.
} \label{map1}
\end{figure}

The DMRG method \cite{dmrg} could in principle overcome these limitations. However, its efficiency with periodic boundary conditions (PBC) is undermined by the structure of the quantum entanglement in a ring geometry, and its application has been limited to the case of spinless fermions\cite{dmrg spinless}, or to small rings.
If we consider a partition of our system into two disjoint parts, the quantum entanglement between the two subdivisions can be quantified by the von Neumann entropy $S$. 
The von Neumann entanglement entropy determines the number of states $m$ necessary to efficiently represent the ground-state of the system using a matrix product state (MPS)\cite{frank1}, $m\sim \exp{(\-S)}$. Since the entanglement entropy for a system with periodic boundary conditions is twice the one for open boundary conditions (OBC)\cite{area law}, a much larger basis is needed to simulate it. Besides this fundamental fact, the structure of the MPS used by the DMRG algorithm as an ansatz to approximate the ground state does not properly account for the entanglement introduced by the closed ends of the ring\cite{frank}. 
This is a problem in transport calculations, since the current through the ring scales inversely with the length and the coupling strength in the problem, and high precision is required to study large systems. Ingenious tricks have been introduced\cite{aef}, but their application to general situations is limited.

\begin{figure}
\epsfig {file=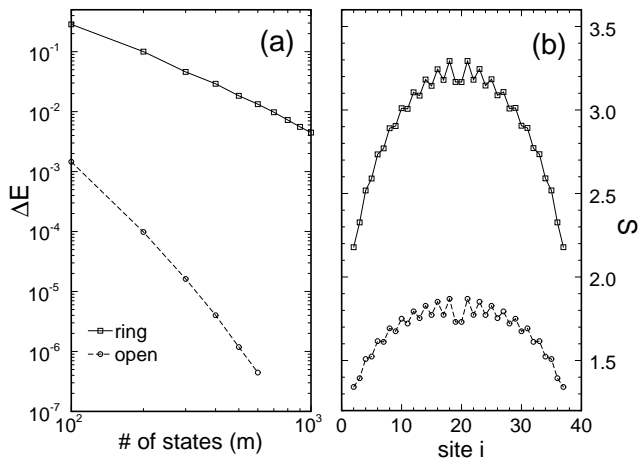,height=85mm,angle=-90}
\caption{
(a) Error in the energy using a ring with PBC, and the equivalent system with OBC, relative to results for OBC using $m=1000$ states.(b) Entanglement entropy for different for different cuts along the chain using both systems. The impurity is in the center. All results are for a system with $L=20$ ($L_{ring}=41$), $J_K=1$, $\phi=\pi/2$, at half-filling.
} \label{convergence}
\end{figure}


Here we revisit a well known canonical transformation that was originally introduced in the context of quantum impurity problems \cite{fisher}, and referred to as a ``folding'' transformation \cite{simon2001}, mapping a Hamiltonian with periodic boundary conditions onto an equivalent model with open boundary conditions. This has two important implications: (i) it reduces the entanglement by half, allowing for a more efficient representation in terms of MPS, and (ii) the OBC enable the DMRG algorithm to simulate this problem optimally.

\section{Models and Method}

\subsection{One-channel Kondo problem}
To illustrate the method, we consider, as an illustration, the case of a side-coupled Kondo impurity connected to a one-dimensional non-interacting chain, as pictured in Fig.1(a). Let us split the chain into a left and right halves, and assume for the moment that the leads have both a finite length $L$. 
The Hamiltonian is written as:
\begin{eqnarray}
H & = & H_\mathrm{leads} + H_\mathrm{boundary} + H_\mathrm{dot} \nonumber \\
H_\mathrm{leads} & = & -t\sum_{\lambda=0,1}\sum_{j=1,\sigma}^{L-1}\left(c_{\lambda,j,\sigma}^\dagger c_{\lambda,j+1,\sigma} + \mathrm{h.c.} \right)\\
H_\mathrm{boundary} & = & -t\sum_{\lambda=0,1}\sum_{\sigma} \left(c^\dagger_{\lambda,1,\sigma}c_{0,\sigma} + \mathrm{h.c.}\right) \nonumber \\
H_\mathrm{dot} & = & J_K \vec{S}\cdotp \vec{s}_0 - hS^z\nonumber 
\label{hamiltonian}
\end{eqnarray}
where $c_{\lambda,j,\sigma}$ is the electron annihilation operator acting on site $j$ of lead $\lambda$ (where the values 0,1 correspond to left and right leads), with spin $\sigma=\{\uparrow,\downarrow\}$, and $\vec{S}$ is the spin operator acting on the impurity/dot. The spin $\vec{s}_0$ on site ``0'' connects the chain to the impurity via an antiferromagnetic exchange parametrized by $J_K$. We have also included a Zeeman field $h$ acting on the impurity spin.
We now introduce a symmetric (+) and antisymmetric (-) combination of operators acting on the left and right leads. This is nothing else but an application of the reflection symmetry, yielding new even(+) and odd(-) operators:
\begin{equation}
c_{\pm,j,\sigma} = \frac{1}{\sqrt{2}}(c_{0,j,\sigma} \pm c_{1,j,\sigma}).
\end{equation}
This is a simple change of basis, with the new operators obeying fermionic anti-commutation rules, that yields a curious and convenient identity. The hopping term in $H_\mathrm{boundary}$ becomes:
\begin{equation}
H'_\mathrm{boundary} = -\sqrt{2}t\sum_{\sigma} \left(c^\dagger_{+,1,\sigma}c_{0,\sigma} + \mathrm{h.c.}\right),
\label{leads-dot1}
\end{equation}
while the hopping term into the (-) leads has cancelled identically. As a consequence, the new Hamiltonian will consist of an impurity coupled to a single lead (+), and a second decoupled lead (-). This means that we can just solve the impurity-lead piece of the system independently, leading to a reduction of the Hilbert space by a power of one-half. This transformation has been used in NRG for decades\cite{bulla}. However, to the best of our knowledge, its application to systems with PBC has been curiously ignored, maybe because the NRG method works for the thermodynamic limit\cite{josephson}.

\begin{figure}
\epsfig {file=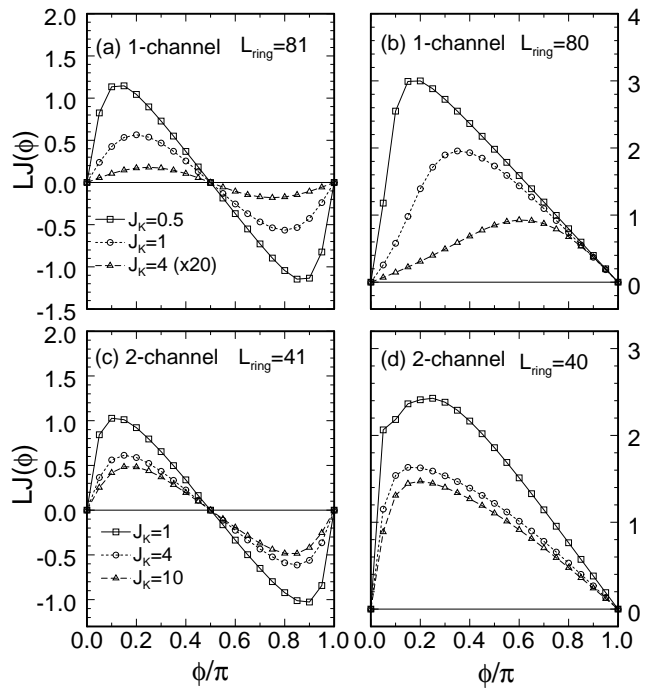,width=85mm}
\caption{
Persistent current as a function of the magnetic phase, for (a)(b) one-channel, and (c)(d) two channel Kondo rings. All simulations are at half-filling. We show results for different values of the Kondo interaction $J_K$ and system sizes.
} \label{current}
\end{figure}

To illustrate the application of the scheme to a case with PBC we studied the behavior of the persistent current at half-filling, as a function of the coupling $J_K$. 
Let us introduce a hopping term connecting the last sites of the two leads, labelled as ``$L$''. 
We can easily see that the hopping term reduces to:
\begin{equation}
H_\mathrm{PBC} = -t \sum_{\sigma} \left(n_{+,L,\sigma}-n_{-,L,\sigma}\right),
\label{boundary1}
\end{equation}
where $n_\pm=c^\dagger_\pm c_\pm$ is the density operator. This term is just a boundary chemical potential with opposite signs for the (+) and (-) leads. But the leads remain decoupled, meaning that we can still solve the problem of a single lead with OBC!

Let us consider now the case of a magnetic flux threading the ring. 
This problem was extensively studied in Ref.\onlinecite{sorensen2005} using DMRG in systems with up to $L_{ring}=35$ sites in an enormous computational effort using PBC.
The flux $\phi$ is introduced by adding a complex phase in the hopping matrix element $t \rightarrow t\exp{(i\phi/L_{ring})}$, where the total length of the ring is $L_{ring}=2L+1$ in our notation. 
By performing a gauge transformation on the fermionic operators, we have the freedom to move the phase to any link along the ring. In particular, we are going to move it to the one connecting the site ``0'' to the leads in $H_\mathrm{boundary}$ in (1). The term containing the phase will now read:
\begin{eqnarray} 
H_\mathrm{boundary} = -t \sum_{\sigma} \left(e^{i\phi}c^\dagger_{0,1,\sigma}c_{0,\sigma} + c^\dagger_{1,1,\sigma}c_{0,\sigma} + \mathrm{h.c.}\right). 
\end{eqnarray}
The flux in the hopping term now introduces a complication: the (-) lead will no longer decouple and, as a consequence, the new, transformed Hamiltonian will have an additional term:
\begin{eqnarray}
H'_\mathrm{boundary} & = & -\sqrt{2}t\sum_{\sigma} \left(e^{i\phi/2}\cos{(\phi/2)}c^\dagger_{+,1,\sigma}c_{0,\sigma} \right. \nonumber \\
& + & \left. ie^{i\phi/2}\sin{(\phi/2)} c^\dagger_{-,1,\sigma}c_{0,\sigma} + \mathrm{h.c.}\right).
\label{leads-dot2}
\end{eqnarray}
This hopping term now couples to both (+) and (-) channels, because we have broken reflection symmetry, but the system still has OBC, as depicted in Fig.1(b).

\begin{figure}
\epsfig {file=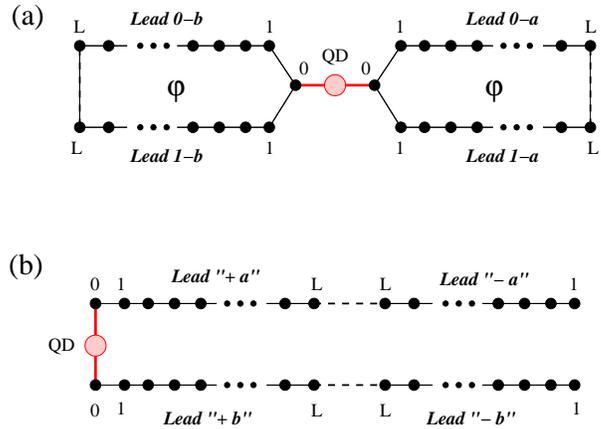,width=90mm}
\caption{
Cartoons for the 2-channel Kondo problem showing (a) the impurity coupled to two rings, and (b) the equivalent system with open boundary conditions.
} \label{map2}
\end{figure}

Fig.\ref{convergence}(a) shows the error in the energy a function of the number of states for a ring with $J_K=1$,$\phi=\pi/2$, $L=20$ ($L_{ring}=41$), using PBC in real space, and the equivalent system with OBC in the transformed basis. As a reference we have used the energy of the system with OBC and keeping $m=1000$ DMRG states.
We clearly see that we can achieve better accuracy with OBC, using a fraction of the number of states. 
This can be explained by looking at the behavior of the entanglement entropy, shown in Fig.\ref{convergence}(b), where the entanglement entropy of the transformed system with OBC is smaller by a factor $\sim2$. In this plot, we show values of $S$ for different cuts along the system, with site ``0'' and the impurity situated both at the center.

Making use of the gauge transformation, we can choose to put the phase factor at the link connecting both leads. In that case, the (-) channel would decouple from the site ``0'', recovering the expression (\ref{leads-dot1}), but the boundary term would have to be corrected:
\begin{eqnarray}
H_\mathrm{PBC} & = & -t \sum_{\sigma} \left(\cos{(\phi)}n_{+,L,\sigma}-\cos{(\phi)}n_{-,L,\sigma} + \right. \nonumber \\ 
& + & \left. i\sin{(\phi)}c^\dagger_{-,L,\sigma}c_{+,L,\sigma} + \mathrm{h.c.} \right).
\label{boundary2}
\end{eqnarray}
This scenario is depicted in Fig.1(c).

The current can be obtained by differentiating the energy as a function of the flux $J=-dE/d\phi$. For numerical convenience, we choose to calculate the expression $J=-it\langle c^\dagger_{0,1,\sigma}c_{0}-c^\dagger_0 c_{0,1,\sigma}\rangle$, since it can be directly obtained from the ground-state. Depending on the gauge choice, we measure it either at the link connect site ``0'', or at the one connecting the two leads. 
In Fig.\ref{current}(a) we show results for a one-channel Kondo ring with $L_{ring}=81$. These results were obtained effortlessly using $m=600$ states, although the error is small enough with $m=200$ for intermediate values of $J_K$.  
The profile of the persistent current can be deduced from symmetry considerations \cite{aligia,simon2001}. Using reflection symmetry around the dot $(c_{0,j,\sigma} \rightarrow c_{1,j,\sigma})$, we find that $J(\phi)=J(-\phi)$. This indicates that it is sufficient to estimate $j$ in the interval $0 \leq \phi \leq \pi$. Furthermore, for the present case with $N=L_{ring}$ odd, an electron-hole transformation on the fermions $(c_{\lambda,j,\sigma} \rightarrow c^\dagger_{\lambda,j,\sigma})$ and also on the impurity, maps $H(-\phi)$ onto $H(\phi+\pi)$ and $J(-\phi) = -J(\phi+\pi)$. Combining both results we obtain that $J(\phi)=J(\phi+\pi)$, and $J(\pi/2) = 0$. This means that the persistent current will have a periodicity in $\pi$, instead of $2\pi$. 

With little additional complication, the transformation can be applied to a ring with an even number of sites. For comparison, we show results for a system with $L_{ring}=80$ in Fig.\ref{current}(b). Using similar symmetry arguments, one can show that the periodicity in this case should be $2\pi$. In all cases the current is rapidly suppressed with increasing $J_K$. For large $J_K$, the size of the ``Kondo cloud'' will be confined to the site ``0'' in direct contact with the impurity. This local tightly bound singlet will suppress conduction through the ring.


\subsection{Two-channel Kondo problem}
We now generalize the above considerations to the case of a two-channel Kondo model\cite{cox}. In this case, the impurity is also coupled to a second channel, represented by a second ring in Fig.\ref{map2}(a). We can perform a similar canonical transformation in both rings independently. We assume that a magnetic flux $\phi$ is threading both rings, and we put the phase terms on the connecting link at the end of each lead as in Eq.(\ref{boundary2}). This open both rings symmetrically about the impurity site (see Fig.\ref{map2}(b)), yielding an equivalent one-dimensional system with OBC, that can easily and efficiently be simulated with the DMRG method.
The remarkable aspect in this case is that we have eliminated {\it two} closed rings from the problem.

\begin{figure}
\epsfig {file=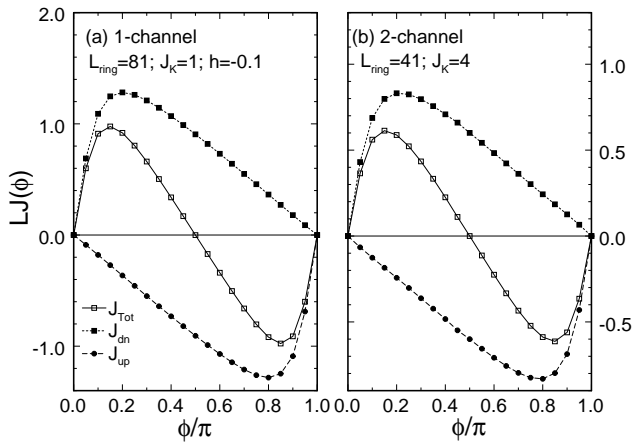,height=83mm,angle=-90}
\caption{
Current per spin for (a) the one-channel Kondo problem with a Zeeman field, and (b) the two-channel Kondo problem.
} \label{spin current}
\end{figure}

In this case, we study the problem in the sector with total $S^z=1/2$, and even total number of conduction electrons.
Unlike the one-channel case, in the present case the impurity will be overscreened, with each channel trying to form a singlet with the impurity. Thus, the impurity is expected to be less efficient at suppressing the current through the leads. 

In Figs.\ref{current}(c) and (d) we show results for the persistent current through an impurity coupled to two channels. The length of each lead, for each channel is $L_{ring}=41$ and $L_{ring}=40$ respectively, meaning that the total lengths of the systems are $L_{tot}=83$ and $L_{tot}=81$ sites. Interestingly, the current has the same behavior as in the one-channel problem with $L_{ring}$ odd, even though in this case, since we have two rings, the total number conduction electrons is always even. For $L_{ring}=41$, one would expect a periodicity in $2\pi$ instead of $\pi$. By looking at the currents for each spin sector in Fig.\ref{spin current}(b), we observe that they both have a periodicity of $2\pi$, but they are shifted by a phase $\pi$. Thus, the total current will have a period $\pi$. 
This resembles the situation encountered in a single side-coupled quantum dot with an applied magnetic field, that was proposed as an efficient spin filter in Refs.\onlinecite{torio,aligia2}. A simple intuitive explanation would be to assume that the screening effect of the second ring on the impurity is seen by the first ring as an effective magnetic field acting on the impurity. To prove this picture, we applied a small Zeeman field to the one-channel problem, and we show the results in Fig.\ref{spin current}(a). We clearly observe the same behavior as in the two-channel case, with curves that are qualitatively indistinguishable.

\section{Summary and Conclusions}
 To summarize, we have applied a ``folding'' canonical transformation to lattice models of impurities coupled to rings, mapping the problems onto equivalent systems with open boundary conditions. As a remarkable counter-intuitive result, we find that a system with open ends can realize persistent current when transformed to the original real-space basis. As a consequence, entanglement is reduced by a factor $2$, and makes the problem suitable for efficient DMRG simulations of large systems with little or moderate effort, as shown here for the cases of the one-channel and two-channel Kondo problems.
The folding transformation 
cannot be applied to disordered rings 
, or to problems with bulk interactions, since many-body terms would introduce long-range interactions between (+) and (-) leads. However, the significant entanglement reduction opens the doors to finite temperature calculations\cite{ancilla}, and the study of complex impurity problems.

\section{Acknowledgments}
We would like thank C. Bolech and A. Aligia for useful discussions. 
AEF is grateful to NSF for funding under grant DMR-0955707.


\begin{thebibliography}{99}

\bibitem{Hewson} A.C. Hewson, {\it The Kondo Problem to Heavy Fermions}, Cambridge Univ. Press (1997)
\bibitem{Hanson2007} R. Hanson, {\it et al.}
, Rev. Mod. Phys. {\bf 79}, 1217 (2007). See also L. P. Kouwenhoven, D. G. Austing, and S. Tarucha, Rep. Prog. Phys {\bf 64}, 701 (2001).
\bibitem{Goldhaber-GordonReview} 
M. Grobis, {\it et al.}
, {\it Handbook of Magnetism and Advanced Magnetic Materials, Vol. 5.}, Wiley (2009).
\bibitem{affleck-sorensen-2006} I. Affleck and E. S. Sorensen, Phys. Rev. B {\bf 75}, 165316 (2007).
\bibitem{carlos} E. V. Anda, C. A. B\"user, G. Chiappe and M.A. Davidovich, \prb {\bf 66}, 035307 (2002).
\bibitem{imry} M. B\"uttiker, Y. Imry and R. Landauer, Phys. Lett. {\bf 96A}, 365 (1983).
\bibitem{rejec} T. Rejec and A. Ramsak,\prb {\bf 68}, 033306 (2003); \prb {\bf 68}, 035342 (2003).
\bibitem{armin} A. Rahmani, C.-Y. Hou, A. Feiguin, C. Chamon, and I. Affleck,
Phys. Rev. Lett. {\bf 105}, 226803 (2010).
\bibitem{shastry} B. S. Shastry and B. Sutherland, Phys. Rev. Lett. {\bf 65}, 243 (1990).
\bibitem{doug} D. J. Scalapino, S. R. White, and S. Zhang, Phys. Rev. B {\bf 47}, 7995 (1993).
\bibitem{rodrigo} R. G. Pereira, N. Laflorencie, I. Affleck, and B. I. Halperin, Phys. Rev. B {\bf 77}, 125327 (2008).
\bibitem{chetan} C. Nayak et al., Rev. Mod. Phys. 80, 1083 (2008).
\bibitem{julian} J. Rinc\'on, A. A. Aligia, and K. Hallberg, Phys. Rev. B {\bf 79}, 035112 (2009).

\bibitem{cho2001} S. Y. Cho, K. Kang, C. K. Kim, and C.-M. Ryu, \prb {\bf 64}, 033314 (2001).
\bibitem{eckle} H.-P. Eckle, H. Johanesson, and C. Stafford, \prl {\bf 87}, 016602 (2001).
\bibitem{aligia} A. A. Aligia, \prb {\bf 66}, 165303 (2002).
\bibitem{josephson} M.-S. Choi, M. Lee, K. Kang, and W. Belzig, Phys. Rev. B {\bf 70}, 020502(R) (2004).
\bibitem{sorensen2005} E. S. Sorensen, and I. Affleck, \prl {\bf 94}, 086601 (2005).
\bibitem{qmc} E. Gull, {\it et al.}, Rev. Mod. Phys. {\bf 83}, 349 (2011).
\bibitem{bulla} R. Bulla, T. A. Costi, and T. Pruschke, Rev. Mod. Phys. {\bf 80}, 395 (2008).
\bibitem{dmrg} S.R. White, \prl {\bf 69}, 2863 (1992); \prb {\bf 48}, 10345 (1993).
\bibitem{frank1} F. Verstraete, M. M. Wolf, D. Perez-Garcia, and J. I. Cirac, Phys. Rev. Lett. {\bf 96}, 220601 (2006); N. Schuch, M. M. Wolf, F. Verstraete, and J. I. Cirac, Phys. Rev. Lett. {\bf 100}, 030504 (2008).
\bibitem{area law} J. Eisert, M. Cramer, and M. B. Plenio, Rev. Mod. Phys. {\bf 82}, 277 (2010).
\bibitem{frank} F. Verstraete, D. Porras, and J. I. Cirac, Phys. Rev. Lett. {\bf 93}, 227205 (2004). See also P. Pippan, S. R. White, and H. G. Evertz, Phys. Rev. B {\bf 81}, 081103 (2010).
\bibitem{aef} A. E. Feiguin, S. R. White, and D. J. Scalapino, Phys. Rev. B {\bf 75}, 024505 (2007). See also A. E. Feiguin, S. R. White, D. J. Scalapino, and I. Affleck, Phys. Rev. Lett. {\bf 101}, 217001 (2008).
\bibitem{fisher} C. L. Kane and M. P. A. Fisher, Phys. Rev. Lett. {\bf 68}, 1220 (1992).
\bibitem{simon2001} P. Simon, and I. Affleck, \prb {\bf 64}, 085308 (2001).
\bibitem{dmrg spinless} V. Meden, U. Schollw\"ock, \prb {\bf 67}, 035106 (2003); \prb{67}, 193303 (2003). See also R. A. Molina {\it et at.}, Eur. Phys. J. B {\bf 39}, 107 (2004).
\bibitem{cox} D. L. Cox, and A. Zawadowski, Advances in Physics, {\bf 47} 599 (1998). See also A. K. Mitchell, D.E. Logan and H.R. Krishnamurthy, arXiv:1103.5038.
\bibitem{torio} M. E. Torio, K. Hallberg, A. H. Ceccatto, and C. R. Proetto,
Phys. Rev. B {\bf 65}, 085302 (2002). M. E. Torio, K. Hallberg, S. Flach, A. E. Miroshnichenko, and M. Titov, Eur. Phys. J. B {\bf 37}, 399 (2004). 
\bibitem{aligia2} A. A. Aligia and L. A. Salguero, Phys. Rev. B 70, 075307 (2004).
\bibitem{ancilla} A. E. Feiguin, and S. R. White, Phys. Rev. B {\bf 72}, 220401R (2005).


\end{thebibliography}
\end{document}